\documentstyle[12pt,psfig,epsf]{article}
\newcommand{\be}{\begin{equation}}
\newcommand{\ee}{\end{equation}}   

\newcommand{\bea}{\begin{eqnarray}}
\newcommand{\eea}{\end{eqnarray}}
\newcounter{dafigcounter}
\def\thedafigcounter{\arabic{dafigcounter}}
\newdimen \figleftportion
\newdimen \figrightportion
\newcommand{\partfig}[4]{\refstepcounter{dafigcounter}
\figrightportion=\textwidth
\figleftportion=#3
\advance \figleftportion by \epsfxsize
\advance \figrightportion by -\figleftportion
\advance \figrightportion by -4mm
\noindent
\begin{minipage}[b]{#3}
\mbox{\epsffile{#1.eps}}
\end{minipage}\hfill
\begin{minipage}[b]{\figrightportion}
\label{#1}Figure \thedafigcounter :\ #2
\vspace*{#4}
\end{minipage}}

\newcommand{\widefig}[2]{\refstepcounter{dafigcounter}
\noindent
\begin{minipage}{\textwidth}
\begin{center}
\mbox{\epsffile{#1.eps}}
\end{center}
\label{#1}Figure \thedafigcounter :\ #2
\end{minipage}\vskip 5mm}

\begin{document}
\title{Microscopic Analysis of Thermodynamic Parameters from 160 MeV/n -
160 GeV/n\footnote{Work supported  
by BMBF, DFG, GSI}
}

\author{M.~Bleicher, S.~A.~Bass, M.~Belkacem, J.~Brachmann,\\ 
M.~Brandstetter, C.~Ernst, L.~Gerland, J.~Konopka, S.~Soff, \\
C.~Spieles, H.~Weber, H.~St\"ocker, W.~Greiner
\\[0.2cm]
{\small Institut f\"ur Theoretische Physik der J.W.Goethe Universit\"at}\\
{\small Postfach 111932, D-60054 Frankfurt a.M., Germany}
\\[0.5cm]}

\maketitle
\begin{abstract}
Microscopic calculations
of central collisions between heavy nuclei are used to study fragment
production and the creation of collective flow. It is shown that the 
final phase space distributions are compatible with the expectations 
from a thermally equilibrated source, which in addition exhibits a collective
transverse expansion. However, the microscopic analyses of the transient
states in the reaction stages of highest density and during the
expansion show that the system does not reach global equilibrium. Even if
a considerable amount of equilibration is assumed, the connection of the 
measurable final state to the macroscopic parameters, e.g.\ the temperature,
of the transient ''equilibrium'' state remains ambiguous.
\end{abstract}

\newpage

\section{Motivation}

Experimental information on yields of fragments and hadrons from high
energy nuclear reactions has long been used to determine the
thermodynamic state of the droplet of hot, dense matter before
decay/freeze-out \cite{fermi50,landau,hagedorn,siemens,mekijan,csernai,
ogloblin}. 

Recently, the spectra of light and medium mass fragments in the energy region
from 150 MeV/n - 2 GeV/n \cite{Kuhn93} and hadrons have also been included
into the description and the large discrepancy between the yield ratios
and the ''thermal'' slopes became apparent.

At much higher energies, 10-15 GeV/n \& 160-200 GeV/n, both, the yields
\cite{satz,heinz} and  spectra \cite{pbm} have been analyzed. The thermal 
parameters (T, $\mu_B$, $\mu_s$), as well as dynamic flow, both in 
longitudinal and transverse
direction, have been obtained. Ambigous conclusions have been reached,
e.g. Braun-Munzinger et al.\cite{pbm} claim to extract temperatures of about
130MeV at AGS energies, while Cleymans et al.\cite{cleym} find only 110MeV.
At 200 GeV/n, Cleymans et al. find $T\approx $190-200 MeV and 
Braun-Munzinger et al. sees 160-170MeV.

Here we present - based on our newly developed microscopic UrQMD model -
a critical discussion of the reliability of the theoretical input and
macroscopic model assumptions used in most of the above analysis:

\section{Hadrons from a Thermal Source}

The idea behind the thermal scenario is as follows: one assumes a 
thermalized system
with a constant density $\rho(r)$ (box profile), a constant temperature
$T(r)$  and a
linear radial and longitudinal flow velocity profile $\beta_{\perp}(r)$, 
$\beta_{||}(r)$.
These parameters are assumed to be the same for all hadrons/fragments. 
At some time
$t^{\rm break-up}$ and density $\rho^{\rm break-up}$, the system
decouples as a whole and the particles are emitted instantaneously 
from the whole volume of the thermal source. A
complete loss of memory results, due to thermalization - the emitted 
particles carry no information about the evolution of the source.

It is of utmost importance to take care of several corrections if one
wants to use
\begin{itemize}
\item{the inverse slope parameter T as thermometer\cite{stoecker},(since
the feeding from $\Delta$'s, as well as the radial flow needs to be
incorporated into the analysis),}

\item{d/p, $\pi$/p resp., as an entropymeter\cite{siemens}(here even
more feeding effects from unstable states has to be considered.}
\end{itemize}

\noindent
The hadron densities from a thermal source are given by
\be
\rho_i = \frac{g_i}{2\pi^2} \int_0^{\infty} {\rm d}p\, 
         \frac{p^2}{\exp[(E_i-\mu_i)/T]\pm 1}\quad .
\ee
This yield is the nascent (primordial) distribution. On top of this rates,
feeding contributions, as mentioned above, from other particles are 
taken into account by
\be
\rho^{\rm final}_{\pi}(\mu,T) = \rho_{\pi}(\mu,T)
                                +\rho_{\Delta}(\mu,T)N^{\pi}(\Delta) 
                                + \dots\quad .
\ee
In addition, the Hagedorn volume correction can be applied to
incorporate changes due to the finite volume of the hadrons. 

It seems that a two parameter fit ($\mu_q$, T, $\mu_s$ is fixed by 
strangeness conservation) to the hadronic freeze-out data describes the
experimental results well, if feeding is included (cf.
Fig.\ref{fs170})\cite{pbm,spieles}. 
Does this compatibility with a thermal source proof volume emission from a globally equilibrated source? 

\epsfxsize = 9.1cm 
\widefig{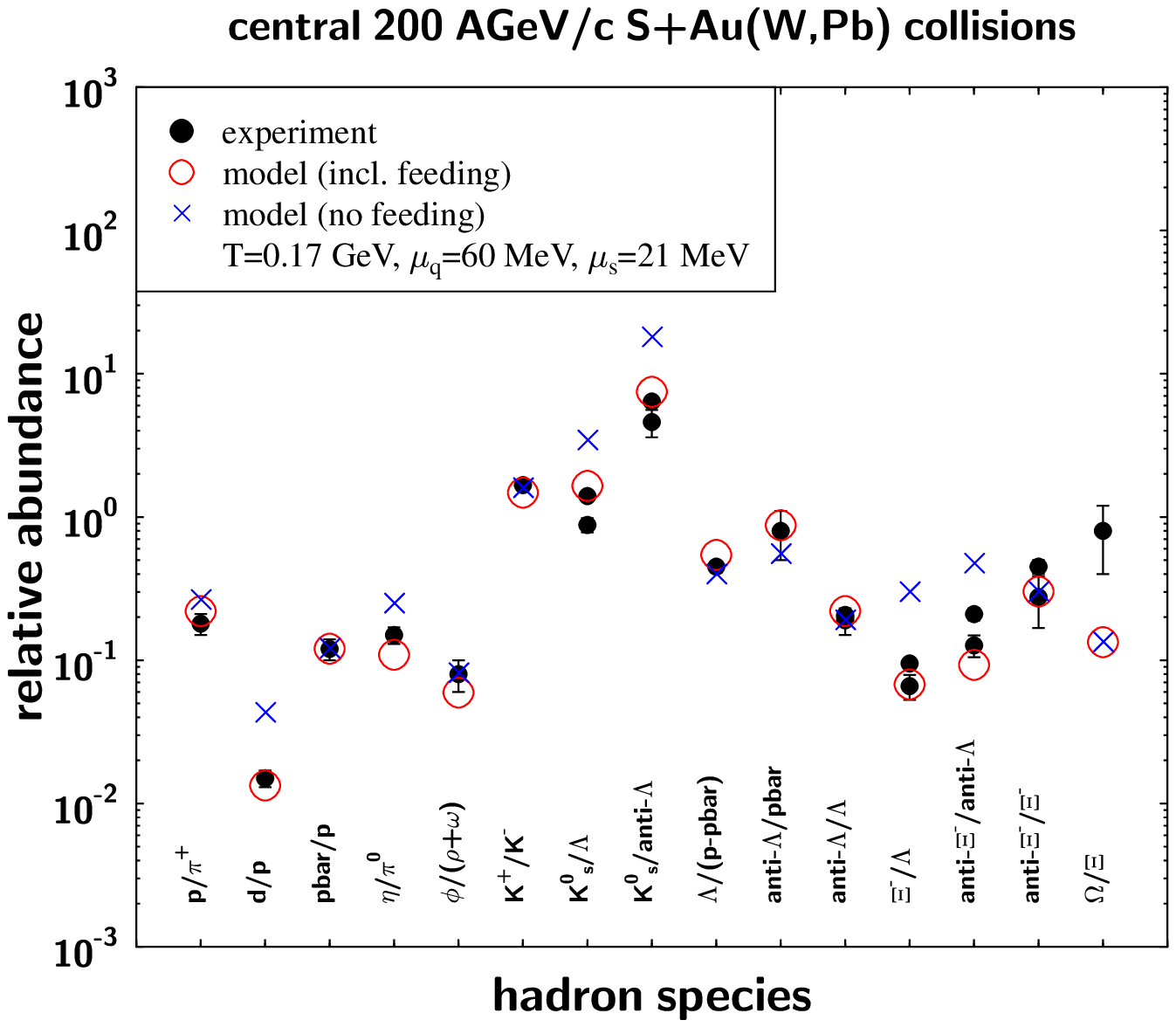}
{Hadron production from a thermal source. $T =170$~MeV,
$\mu_q=60$~MeV, $\mu_s=21$~MeV. Total strangeness is 
conserved\cite{pbm,spieles}.}

\noindent
The same quality of fit to the hadron yields is found if the evolution from 
an initial quark gluon plasma via a $1^{\rm st}$order phase transition to 
hadronic matter is assumed, even with non-equilibrium evaporation of particles during the time evolution of the system\cite{spieles}.

\section{Microscopic Model - Macroscopic Analysis}

Let us start by calculating some macroscopic quatities within the
framework of a microscopic model. 
In order to get information on the thermal collective energy sharing,   
we restrict our analysis for the moment to the transverse momentum
spectra only. This procedure is motivated by the fact that transverse momenta
are newly created and are not directly connected to the initial
(longitudinal)
beam momentum. A possible collective expansion in longitudinal direction
cannot unambiguously be related to the properties of the hot and dense
reaction zone because the system may have memory about its history,
in particular the incident momentum of the beam.

The collective expansion velocities exhibit a decrease with
increasing hadron/fragment mass, which is displayed (for 150 AMeV) in
Fig.\ \ref{wilder_tp0fit}.
The results on the averaged collective transverse momentum and
temperature
from the QMD calculation (full symbols) compare reasonably well to a
experimental compilation of available data on central Au+Au
collisions~\cite{Cof94}, which have been analysed with the very same
method.
Similar collective flow velocities and temperatures have also been
reported from the EOS collaboration~\cite{Lisa94}.

\epsfxsize = 14cm 
\widefig{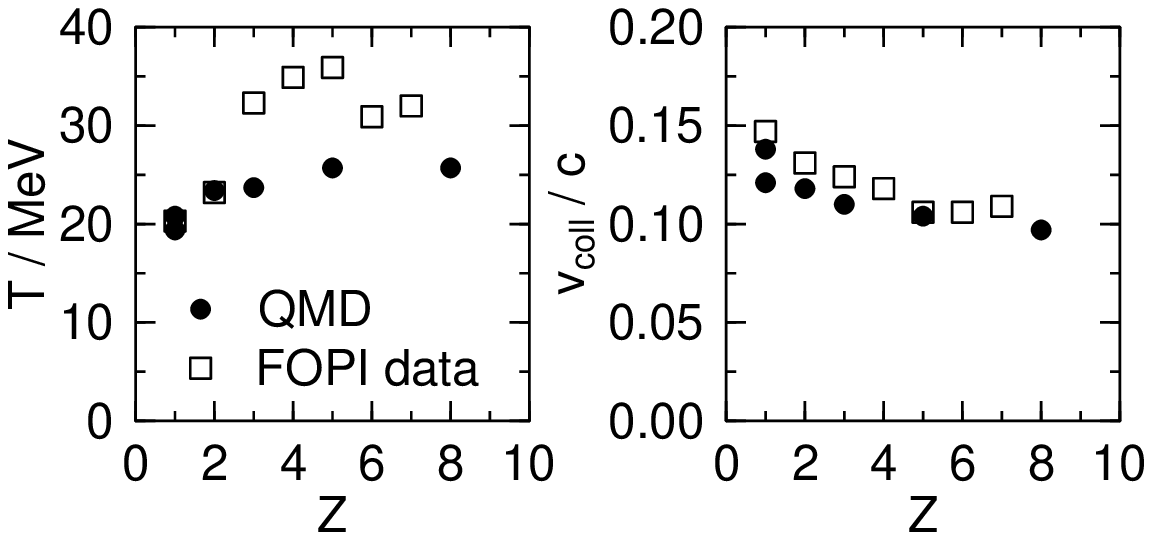}{
Temperature and flow velocities for fragments. The QMD results (closed circles)
deduced from the fits to the transverse momentum spectra in Fig.\
\protect{\ref{wilder_pt}} are compared to data
of the FOPI-collaboration \protect{\cite{Cof94}} (open squares).}

Fig.\ \ref{wilder_pt}
shows transverse momentum spectra of various charged fragments
obtained with QMD for the system Au (150 MeV/nucleon, b=0) + Au
(symbols) together with fits to these calculated data, which are based
on the
assumptions from above. In fact, the corresponding count rates have been
fitted, rather than the invariant distributions, which are displayed. 
All spectra are compatible with temperatures
between 20 and 25 MeV and averaged collective flow velocities of
0.1--0.13 c.

On the contrary
these rather high temperatures are in variance with the expectations
($T\approx 8$ MeV at 150 MeV/nucleon) from a quantumstatistical analysis
of the large number of intermediate mass fragments observed in the very
same system~\cite{Kuhn93}. This seems to support the idea of collective
radial flow in addition. To illustrate this radial flow we take a look
at the coordinate space distribution of the expanding fireball. Fig.\ \ref{wildert50}
shows a typical snapshot of the expansion phase of a heavy ion reaction.
Density contours are shown together with local collective expansion
velocities. A strong correlation between configuration and velocity space is
observed. Inside any of these cells the velocity distribution exhibits a gaussian 
shape. 
\vspace*{0.5cm}

\epsfxsize = 7cm 
\partfig{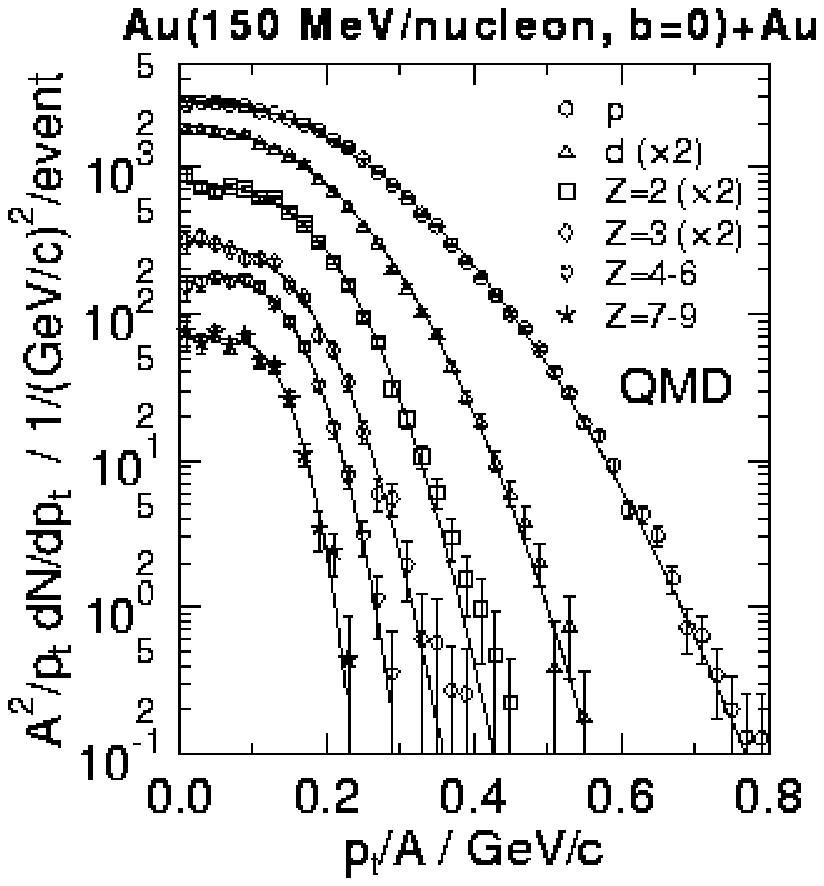}
{Invariant transverse velocity spectra of various
reaction products of Au (150 MeV/nucleon) + Au at vanishing impact parameter.
The predictions of the Quantum Molecular Dynamics model (symbols) have
been fitted with a thermally equillibrated source, which expands azimuthally
symmetric towards the transverse direction (lines). Note, all spectra are    
compatible with temperatures between 20 and 25 MeV and an averaged transverse
collective velocity of 0.11--0.13 c.}{0.3cm}{0.3cm}

\epsfxsize = 7cm 
\partfig{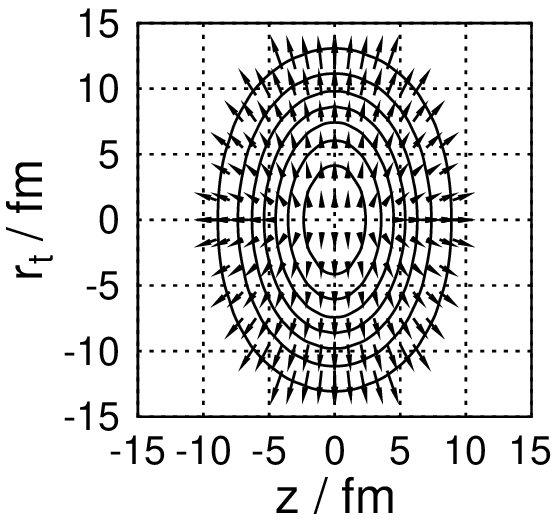}
{Snapshot of the expansion after 50 fm/c starting from a 5 fm separation
of projectile and target surfaces. Density contours are at 0.1--0.7   
$\rho_0$. The arrows indicate the direction and the absolute velocity
(proportional to the length) of the collective motion of the individual   
fluid cells.}{0.4cm}{2cm}

These ``local temperatures'' vary only slightly over large
volumes in real space which supports
a common analysis of a larger number of cells.
For this purpose the central reaction volume is defined as the sum of
all cells in which the local density exceeds half the maximum density at
this instant, implying a time-dependent volume.
In Fig.\ \ref{wilder_times} the properties of the excited
nuclear matter inside this volume are displayed as a function of time.
Even in the late stages of the expansion this zone still contains
$\approx 1/3$ of the mass of the entire Au+Au system. The evolution of
the spreads of the local velocity distributions suggests a rapid
cooling,
which goes in line with a strong density decrease. The different
temperatures tend to converge in the late stages only.

At densities around 0.1--0.5 $\varrho_0$, where the freeze-out of
fragments is supposed to happen, the corresponding temperatures have 
dropped below 10 MeV. This is no longer in vast disagreement with the 
chemical temperatures needed for the understanding of the large intermediate
mass fragment multiplicities~\cite{Kuhn93}. 
\vspace*{0.5cm}

\epsfxsize = 5cm 
\partfig{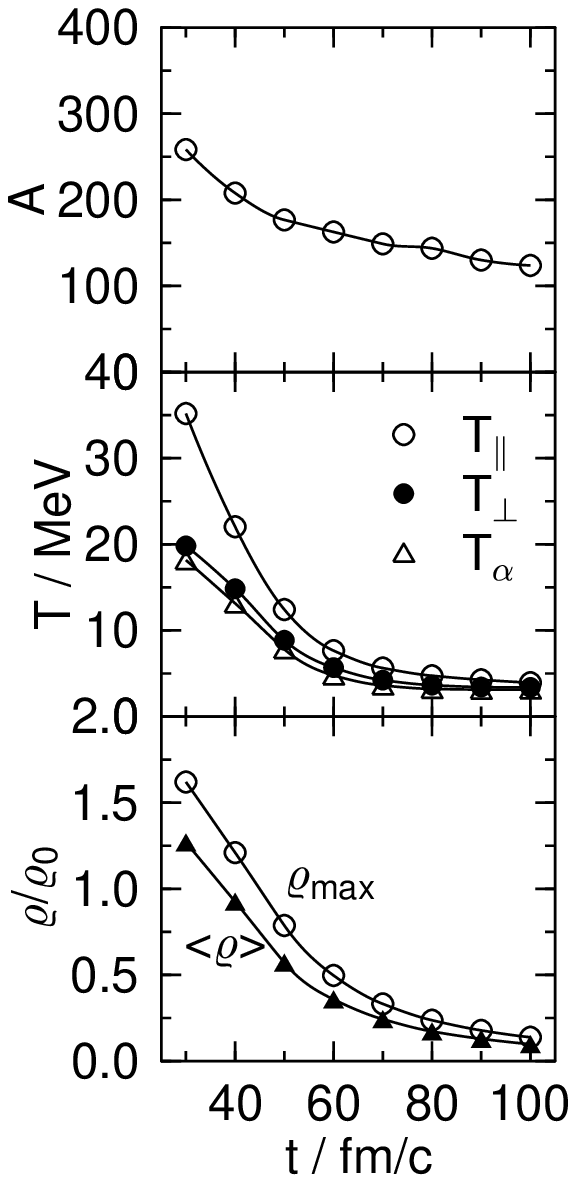}
{Thermodynamics in the central reaction zone, i.e.\ the volume
where the local density is at least half of the maximum value. Mass
content a), spreads of the local velocity distributions b), and maximum
as well as averaged density c) are displayed as a function of time.}   
{0.5cm}{6.5cm}

\section{Analysing the Source Microscopically}

We have shown in the two previous sections that the final state spectra,
although they seem to be perfectly described by an equilibrated, expanding
source, suggest too high temperatures even if collective flow effects are taken
into account. Locally much lower temperatures are observed in the very same
class of events.

In turn the question arises what is wrong with the assumptions which
underlied the global fit. Let us analyse the density $\rho(r)$, and the radial flow at
''break-up time'' $t\approx 50$~fm/c. From Fig.
\ref{wilder_assump} we conclude that  
\begin{itemize}
\item{the radial flow rises linearly with the transverse distance $\beta(r) \approx a\cdot r$, as assumed in many simplified macroscopic
models,
}
\item{the temperature $T$ is roughly constant over the radius at freeze-out (not
shown in the Figure)\cite{konopka},
}
\item{the density distribution $\rho(r)$ falls of smoothly. This, however is in strong contrast to
the box profiles used in the macroscopic analysis!
}
\end{itemize}

\noindent
How can we understand this difference? We know from textbooks that
$PV=NRT$. Dividing by $V$ (using $R=1$) we get 
\be
P=\rho T \quad.
\ee 
However, in hydrodynamics the relation 
\be
\frac{\rm d}{{\rm d}t}\vec{p}=-\nabla P(\rho,T)
\ee
holds. This means, the gradients of the pressure P lead to flow, starting with the (isentropic)
expansion into the vacuum at the surface. Even if there were an 
initial box profile, it is
changed into a complex final shape $\rho(r)$.

\epsfxsize = 14cm 
\widefig{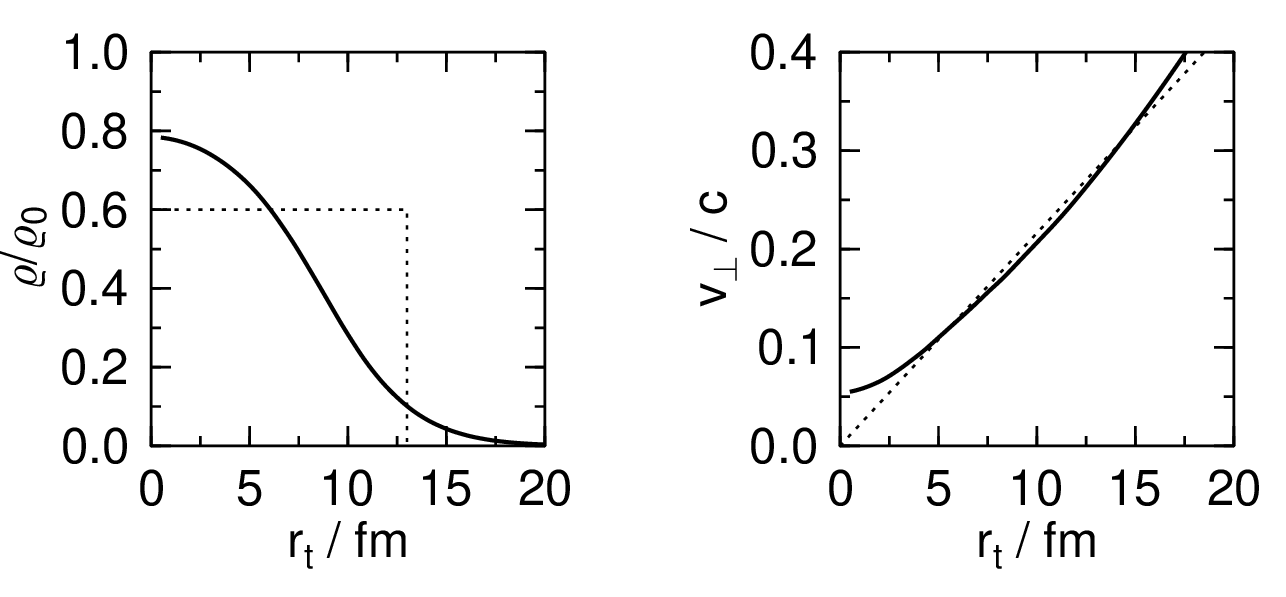}
{Local density and collective velocities as a function of the
transverse distance in the $z=0$ plane. The Figure corresponds again
to the situation after 50 fm/c. The calculated shapes (solid lines) are
compared to the assumptions, which entered the global fit (dotted lines).
The longitudinal as well as the azimuthal collective velocity vanish.
$v_\perp$ does not reach 0 for $r_\perp\rightarrow 0$ due to the fact  
that the innermost cell is not at $r_\perp = 0$.}

This may not be considered a drastic source of uncertainty. However, 
 the problem comes from the assumption of a 
$\beta^{\rm max}$ at $R^{\rm break-up}$. As expressed in Fig.\
\ref{wilder_assump}, the matter distribution has a long 
tail, which picks up high flow velocity components. In the box model
such ''cells'' do not exist. Their contribution to the high energy tails
of the spectra must then be ''simulated'' by an artificially enhanced
''temperature'' in the fit. The microscopic analysis suggests, on the contrary, the cells to be considerably cooler. The
specific combinations of densities and temperatures, which are traversed in
the course of the reaction, are important. The complete space-time 
evolution is needed
to study the agreement with the expectations from the quantum statistical
analysis of the fragment distributions in the final state~\cite{Kuhn93}.

%

\section{Thermalisation at AGS/SPS ?}

How does this picture change at higher beam energies, say 10-15 GeV/n, 
resp. 160-200 GeV/n?
Here we also find late clustering of fragments, while high energy
protons cool the reaction zone. Mattiello\cite{raffi2}, as well as 
Bleicher\cite{bleicher} have shown, that the break-up\footnote{In high energy
cascade calculations, the particle freeze-out is defined for each particle individually as the last point 
of strong interaction of that particle.} density distributions $\rho(r)$ are
 not at all box like. Even a supression in $\rho(r)$ for $r\rightarrow 0$~fm was reported \cite{raffi2,bleicher}.

Fig.\ref{rt} shows the transverse velocity profiles of protons and
tritons at freeze-out (top) and the transverse freeze-out densities of
p's, d's, t's and He's (bottom) in central Au+Au (Pb+Pb, resp.) events
at central rapidities.

In general, flow correlations arise for both p's and t's both at AGS 
to SPS 
(Fig.\ref{rt}, top). The radial velocity profiles are convex.
The shapes of the profiles are nearly independent of beam energy and
particle type (cluster). Note, however that unlike at low energies the 
radial velocities level off at about $0.6$c. They are not longer linear 
in $r_\perp$! The 
radial density profiles (Fig.\ref{rt}, bottom) indicate, a wide range of 
distances $r_\perp$ with high $\beta_\perp$ values which contribute 
to the proton distribution. The expansion of the exploding hot hadronic matter
is thus visible by the large apparent ''temperatures'' of the baryon spectra.

In contrast to the protons, there is a strong localisation of the cluster
emission region to the surface of the reaction zone ($r_t\approx 6$~fm).
Cluster formation is attenuated
for $r_t\rightarrow 0$~fm, due to the highly uncorrelated momenta of the
coalescing baryons and the small freeze-out probabilty due to further
collisions with newly produced hadrons. For $r_t>6$~fm cluster
formation is also suppressed, as the emission volume increases and
density descreases,
which leads to larger average distances between the baryons, thus decreasing
the coalescence probability. Even the upcoming momentum correlations
(i.e. radial flow) are not able to counter-balance this effect.

\epsfxsize = 14cm 
\widefig{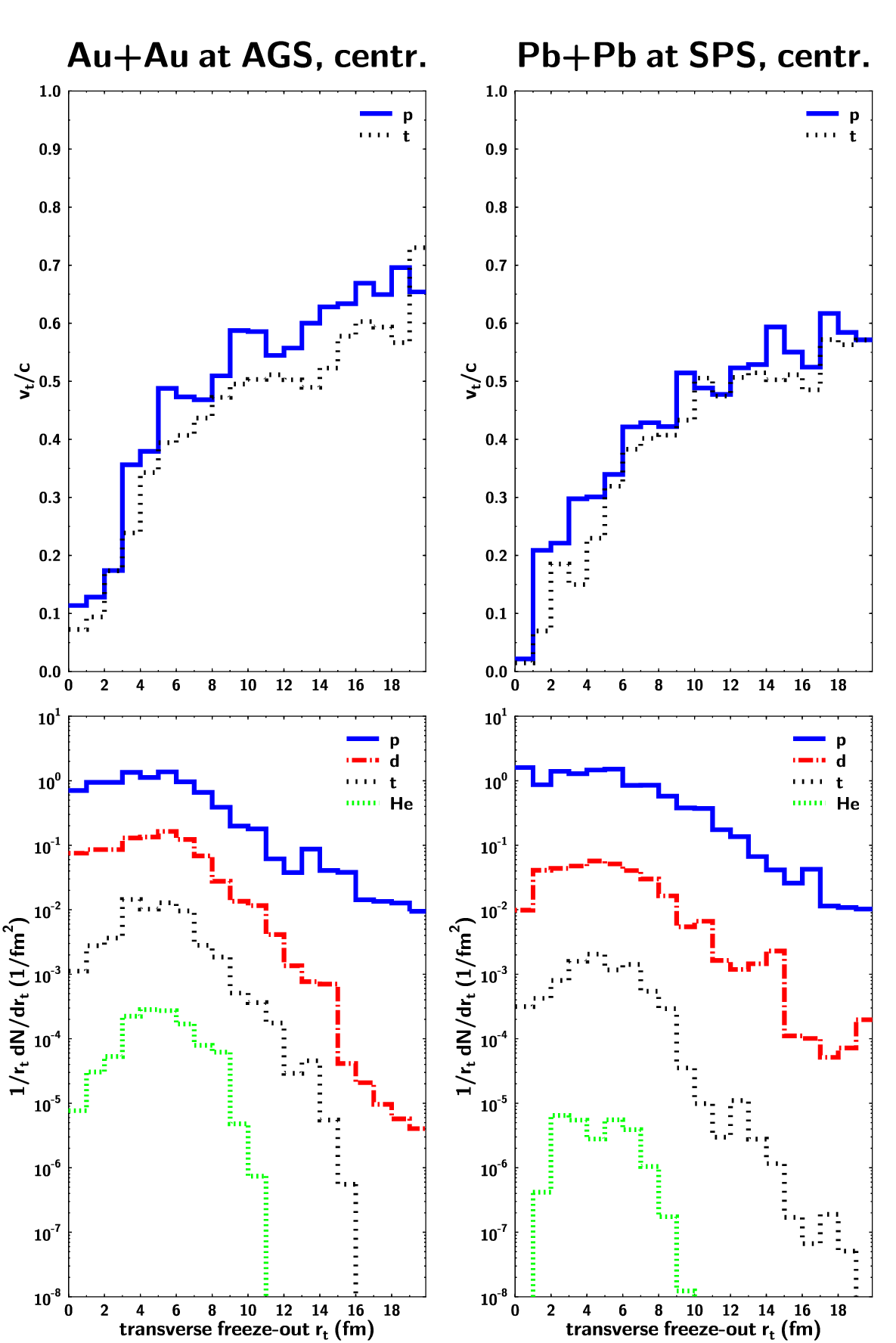}
{Left: Top: The reaction Au+Au at 10.7 AGeV, (central events and
central rapidities selected). Transverse velocity profile of protons and
tritons vs transverse freeze-out radius. Bottom: Radial freeze-out
density of p, d, t, He. Right: The same for Pb+Pb at 160 AGeV.}

In Fig.\ref{mesons} the proton (dotted) freeze-out radius
and time is compared to the freeze-out of $\pi$'s (dashed) and kaons
(dashed dotted, grey). The transverse freeze-out radius of the kaons is
found smaller than those of the pions and protons. The freeze-out time of
the kaons is much earlier than for the $\pi$'s and protons. Most
probable freeze-out times (in fm/c) for K:$\pi$:p are 9:15:23,
which reflects the vastely different cross sections of these particles in 
the nuclear medium ($\sigma \approx$16, 25, 40 mb) . The emission times and the 
positions of different particles are clearly
distinguishable, in contrast to the simple volume breakup of the expanding 
fireball were one  
assumes a equal time freeze-out of all particles. 

\epsfxsize = 14cm 
\widefig{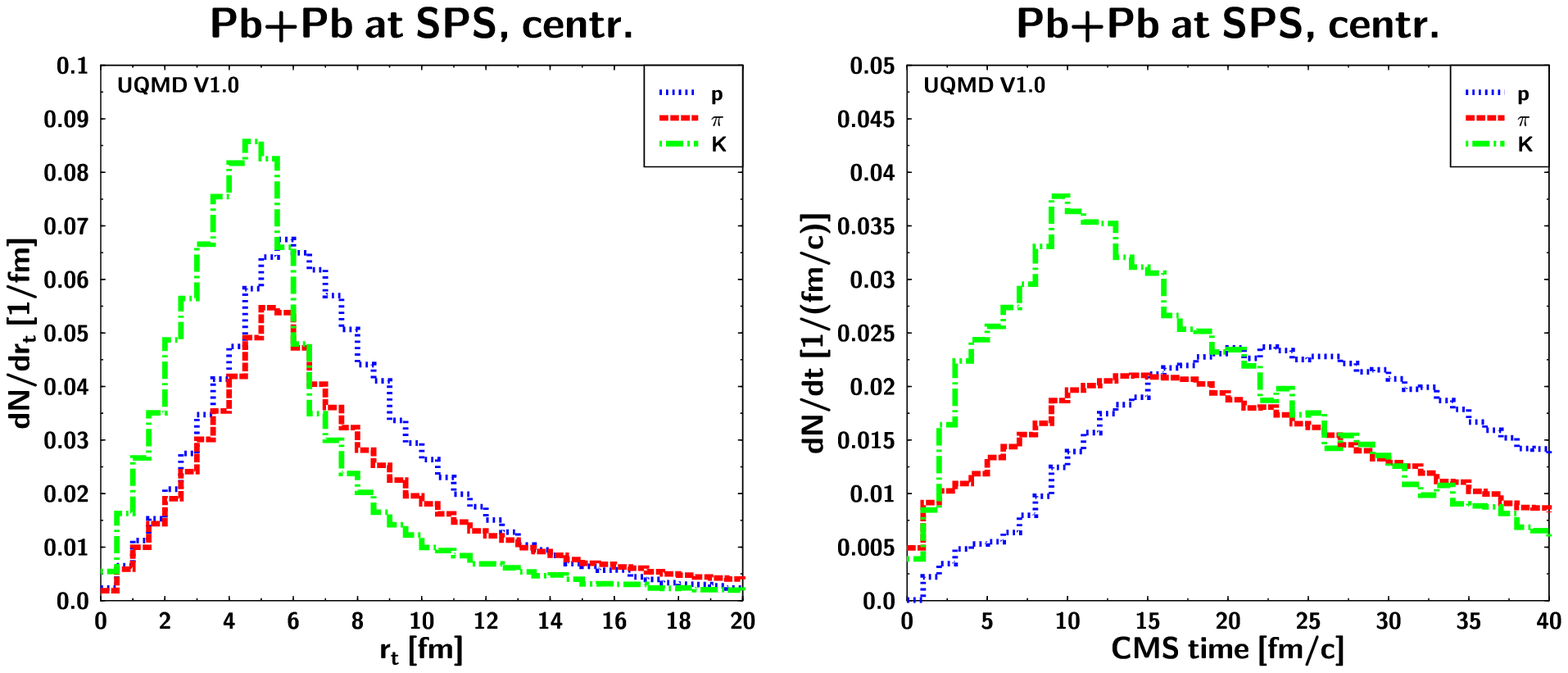}
{Left: Normalized radial freeze-out densities of kaons, pions and
protons in the reaction Pb+Pb at 160 AGeV, centrally triggered. Right:
Freeze-out times of kaons, pions and protons ins the same reaction.} 

Switching off particle decay one finds smaller radii and earlier mean
freeze-out times, for baryons as well as for mesons. The investigated     
freeze-out distribution can be checked via HBT analysis. The radial velocity
profile rises linearly to 0.5c and saturates.

\section{Conclusion}

We have presented UrQMD/QMD transport calculations for head on collisions of
Au+Au (Pb+Pb). Temperatures were deduced by fitting final state particle spectra as
well by analysis of local velocity distributions in the intermediate reaction
stages. The latter are in agreement with the temperatures which are
necessary to explain the large fragment abundances in terms of a
chemical equilibrium. 

In summary, any temperature obtained from particle spectra -- even if
collective flow has been taken into account -- can only serve as a 
rough estimate for the complex space-time conditions at the individual 
freeze-out. In addition we found
that the transverse flow varies for different particles and fragments.
The freeze-out times and densities are not unique -  the configuration space
distributions are essentially unknown from experimental data.

\end{document}